\documentclass[aps,amsfonts,prb]{revtex4}

\usepackage{color}
\usepackage{textcomp}
\usepackage{graphicx}
\usepackage{amssymb}
\usepackage{amsmath}
\usepackage{subfigure}
\usepackage{bm}
\usepackage{amsmath, amsthm, amssymb}
\usepackage{multirow}

\def\beq{\begin{equation}}
\def\eeq{\end{equation}}
\def\bea{\begin{eqnarray}}
\def\eea{\end{eqnarray}}

\begin{document}

\title{Many-body localization phase transition: A simplified strong-randomness approximate renormalization group}

\author{Liangsheng Zhang}
\affiliation{Department of Physics, Princeton University, Princeton, New Jersey 08544, USA}

\author{Bo Zhao}
\affiliation{Department of Physics, Princeton University, Princeton, New Jersey 08544, USA}

\author{Trithep Devakul}
\affiliation{Department of Physics, Princeton University, Princeton, New Jersey 08544, USA}

\author{David A. Huse}
\affiliation{Department of Physics, Princeton University, New Jersey 08544, USA, and Institute for Advanced Study, Princeton, New Jersey 08540, USA}

\begin{abstract}

We present a simplified strong-randomness renormalization group (RG) that captures some aspects of the many-body localization (MBL) phase transition in generic disordered one-dimensional systems.  This RG can be formulated analytically and is mathematically equivalent to a domain coarsening model that has been previously solved.  The critical fixed point distribution and critical exponents (that satisfy the Chayes inequality) are thus obtained analytically or to numerical precision.  This reproduces some, but not all, of the qualitative features of the MBL phase transition that are indicated by previous numerical work and approximate RG studies: our RG might serve as a ``zeroth-order'' approximation for future RG studies.  One interesting feature that we highlight is that the rare Griffiths regions are fractal.  For thermal Griffiths regions within the MBL phase, this feature might be qualitatively correctly captured by our RG.  If this is correct beyond our approximations, then these Griffiths effects are stronger than has been previously assumed.

\end{abstract}

\maketitle

\section{Introduction}

The dynamical quantum phase transition between many-body localization (MBL) and thermalization appears to be a completely new type of quantum phase transition \cite{baa,oh,mg,ph,kbp,blr,nh,av,VHA,ppv,spax,agkmd,lfa,santos,mbmott,ds,bz,goold,mobilityedge,SM,clo,griff}.
This phase transition occurs in the thermodynamic limit of large systems for certain closed quantum many-body systems with quenched disorder.
It separates the {\it thermal phase} where the closed system serves as a ``bath'' for itself and at long times approaches thermal equilibrium at a nonzero temperature and thus a state described by equilibrium quantum statistical mechanics, from the {\it MBL phase} where these statements are not true and instead the system remains localized near its initial state.
Thus it is a phase transition where the long-time behavior of the system stops being given by equilibrium quantum statistical mechanics.
It is also an ``eigenstate phase transition''\cite{hnops}, where the nature of the eigenstates of the system's dynamics changes from thermal and volume-law entangled, to localized and boundary-law entangled.

Many questions remain unanswered about this MBL phase transition.
One theoretical tool that has been highly successful in understanding more ``traditional'' classical and quantum phase transitions and critical systems is the renormalization group (RG).  Two recent papers have formulated and studied approximate RG treatments of the MBL phase transition in one-dimensional systems \cite{VHA,ppv}.
In the present paper, we simplify these RG approaches even more, allowing a more exact study of the resulting fixed point and phase transition within our simplified RG \cite{bz}.
These approximate RGs can serve at least two purposes: (i) as examples that suggest possible properties of these phases and this phase transition, and (ii) as first steps towards possibly developing a more systematic RG treatment of these systems.

One feature of our RG that we highlight, since it was not emphasized in the previous RG studies\cite{VHA,ppv} nor in the recent study of Griffiths effects in the MBL phase\cite{mbmott}, is the possibility that the thermal Griffiths regions within the MBL phase have a fractal dimension $d_f<1$ in these one-dimensional systems.
This is a clear result of our RG, and the mechanism by which this happens seems like it might be more generally valid and not just an artifact of our approximations.
One consequence of such fractal Griffiths regions is that averaged correlations and entanglement within eigenstates in the MBL phase can decay with distance $x$ as stretched exponentials, $\sim\exp{(-(x/x_0)^{d_f})}$, instead of the simple exponentials that one might naively expect.

\section{an approximate rg}\label{next}

Now we present our approximate RG, pointing out and discussing the various assumptions and approximations that are used.
We refer to the two previous RG studies as VHA\cite{VHA} and PVP\cite{ppv}.
Like them, we consider a one-dimensional system whose dynamics is given by a local Hamiltonian, or more generally a local Floquet operator, with quenched randomness.
The system has a MBL phase transition and the system's parameters are near this critical point.
We assume, as in VHA, that each local region of this system can be classified either as thermalizing (T) or insulating (I).
This need not be true of the system at the microscopic scale, but we are assuming that under coarse graining the critical point does flow to an infinite-randomness fixed point where such a ``black and white'' description is correct, and we have already coarse grained enough for this to be a good approximation.
This does happen within our RG (as well in both VHA and PVP), so this assumption is at least internally consistent.
One question for future investigation is whether all these approximate RGs are possibly missing some important physics of the transition by not allowing for some intermediate local behaviors between fully thermalizing and fully insulating to be relevant at the critical fixed point.

If we have such a one-dimensional system and, as we assume, all regions of it can be classified as either T or I, then the system is a chain of ``blocks'' of various lengths $\ell$ that alternate along the chain between T and I.
In the previous RG studies \cite{VHA,ppv}, each such block was characterized by only two parameters: the typical many-body energy level spacing of the block, and some rate at which entanglement can spread from one end of the block to the other end.
Our RG is even more ``simplified'' and characterizes each block only by whether it is T or I and by its many-body level spacing.
The justifications for making this approximation of ignoring the precise value of the ``entanglement rate'' for each block are as follows. For almost all T blocks the entanglement rate at large $\ell$ is much larger than the level spacing which is exponentially small with $\ell$.  As a consequence it is a reasonable approximation to ignore the precise magnitudes of these large T-block entanglement rates and just assume they are very fast compared to the many-body level spacing \cite{bz}.  Near the critical point the I blocks at large $\ell$ are almost all near critical and have entanglement rates whose logarithms are close to that of the level spacing, so we make the approximation that the entanglement rate and the many-body level spacing are equal for each I block \cite{bz}.
These are certainly oversimplifications, and we know that our RG does get some of the physics incorrect, as we discuss below.
The virtues of our RG are its simplicity and that even with this simplicity it does appear to get some of the physics of the MBL transition qualitatively correct.

The many body level spacing of a one-dimensional block of length $\ell$ is $\sim\exp{(-s\ell)}$, where $s$ is the entropy per unit length (e.g., $s=\log{2}$ for a spin-1/2 chain at infinite temperature).
We use the block's length $\ell$ as the parameter that quantifies its level spacing.
The $n$th block in the chain has length $\ell_n$.
Once we have coarse-grained to a scale where adjacent blocks typically differ substantially in length, we can use a strong-randomness RG approach, justified by the typical ratio of many-body level spacings in two adjacent blocks being $\sim\exp{(-s|\ell_n-\ell_{n+1}|)}\ll 1$.  Note that our RG, like the previous ones \cite{VHA,ppv}, assumes that the dynamic critical scaling is given by the many-body level spacing, which is consistent with numerical studies on a spin-chain model's dynamics near its MBL transition \cite{ph}.

\begin{figure}[!htbp]
	\subfigure[TIT rule]{
		\label{fig:TIT_rule} 
		\begin{minipage}[b]{0.49\textwidth}
			\centering
			\includegraphics[width=0.9\textwidth]{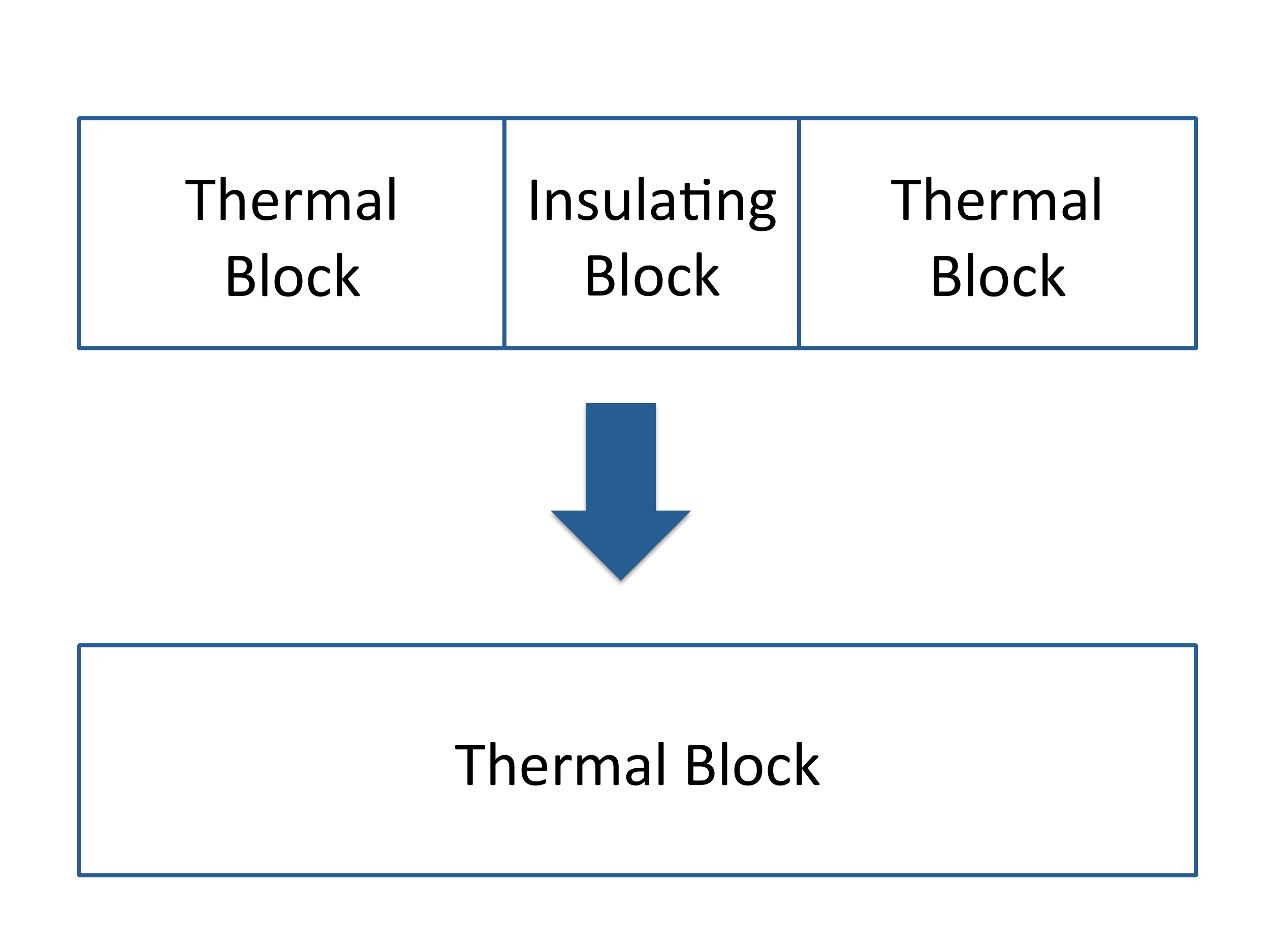}
		\end{minipage}}
	\subfigure[ITI rule]{
		\label{fig:ITI_rule} 
		\begin{minipage}[b]{0.49\textwidth}
			\centering
			\includegraphics[width=0.9\textwidth]{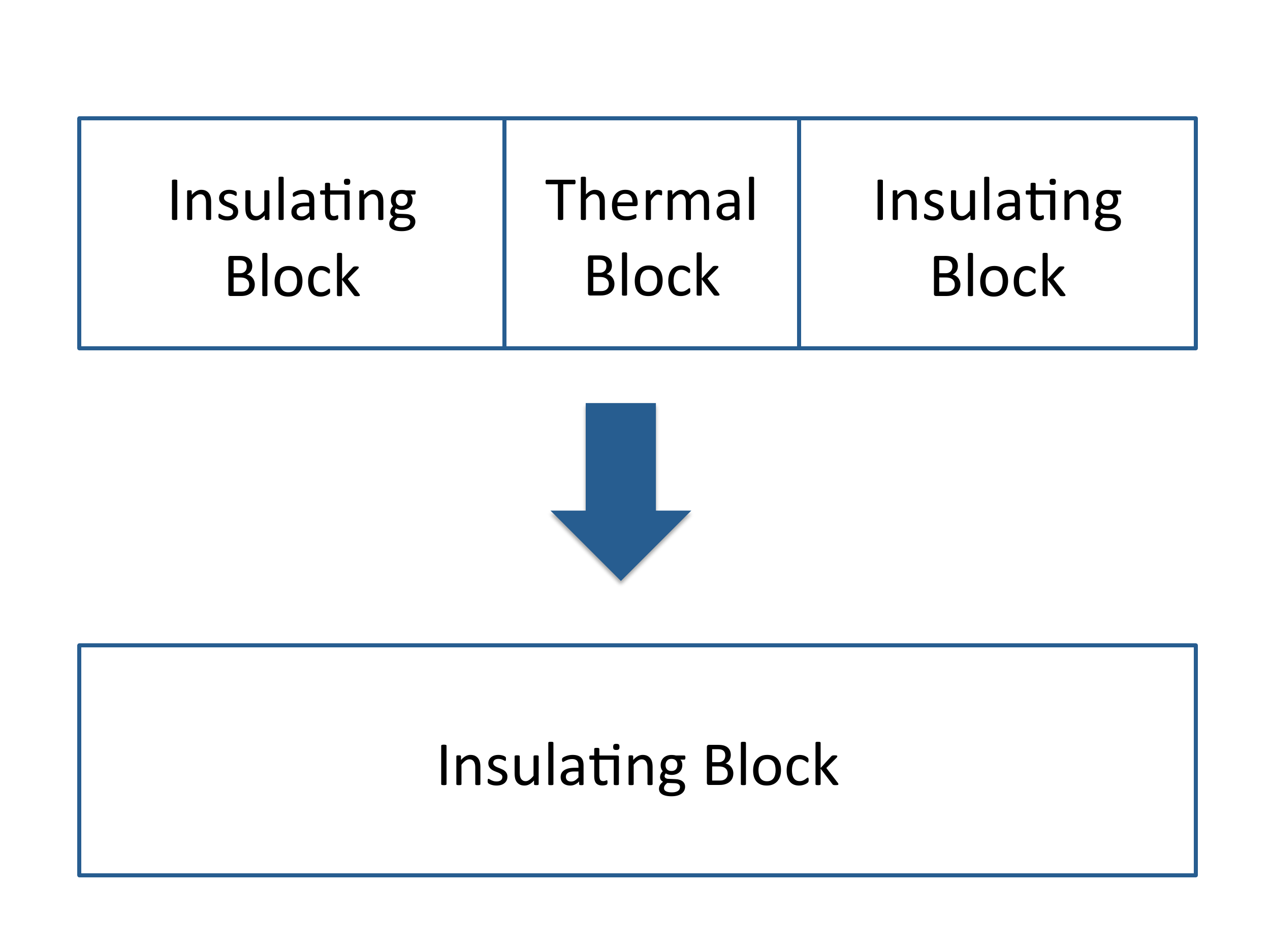}
		\end{minipage}}
	\caption{A sketch of the RG rules. Panel (a) shows the TIT rule where a central I (insulating) block is surrounded by two longer T (thermal) blocks,
and after merging the new longer block is thermal. Panel (b) shows the ITI move where a T block is surrounded by two longer I blocks and the resulting
new block is insulating. }
	\label{fig:RG_rules} 
\end{figure}

A single RG step is simply as follows: Find the shortest remaining block; say it is block $n$.
This is the remaining block with the largest level spacing.
Merge this block with its neighbor blocks on both sides to make a new larger block with length
\beq\label{eq1}
\ell_{new}=\ell_{n-1}+\ell_n+\ell_{n+1} ~.
\eeq
If the shortest block $n$ is an I block, then the new longer block is a T block.
A sketch of this ``TIT'' move is shown in Fig.~\ref{fig:TIT_rule}.
The neighbor blocks, both of which are T blocks that are longer than I block $n$, serve as local ``baths'' and thermalize the shorter central I block.
This relies on the level spacings in both of the longer neighbor T blocks being much smaller than the rate of entanglement spread across the shorter central I block, so the two T blocks get strongly entangled across the I block.
Thus it is quite plausible that the eigenstates of this new T block of length $\ell_{new}$, isolated from its neighbors, are thermal, with thermal ``volume-law'' entanglement within this new block.
In this case, where two longer T blocks thermalize a shorter I block, the approximations we make in our simple RG seem plausible near the MBL phase transition.
It is for the other case, when block $n$ is a T block, that we have to make some questionable assumptions to produce a simple RG.

If the shortest block $n$ is a T block, then the two neighbor blocks are I blocks, and we assume the new longer block that is then made in this RG step is an I block.
Figure \ref{fig:ITI_rule} illustrates this ``ITI'' rule.
The two longer I blocks are isolating the shorter central T block from other T blocks and thus localizing it.
On one level this seems sensible: the rate of entanglement spread across the neighbor I blocks is much smaller than the level spacing of the shorter central T block, so should not mix the eigenstates of the central block.
But if we ask about the rate of entanglement spread across the new long I block of length $\ell_{new}$, a reasonable estimate would be $\sim\exp{(-s(\ell_{n-1}+\ell_{n+1}))}$, since the spread across the central T block is assumed to be very fast.
This suggests that our new long I block may not really be insulating, since this rate is much larger than its level spacing, but we will proceed with the assumption that this new block is indeed I.

One feature of this simple RG that makes it simple is that it is invariant on swapping I and T.
This is because we are treating the process of two T blocks thermalizing a shorter central I block as mathematically the same as two I blocks localizing a shorter central T block.
Unfortunately, we know that the real system does not have such a symmetry.
Facilitating entanglement is very different from, and much ``easier'' than, blocking the spread of entanglement, because interacting quantum systems generically do get entangled.  In the ``battle'' between thermalization and localization, thermalization always has ``the upper hand.''
And we know from all the numerical studies of one-dimensional models that the properties of the phase transition are very asymmetric between the MBL phase and the thermal phase, with the critical point appearing very localized and the changes in systems' properties happening almost all on the thermal side of the phase transition (see, e.g., Refs. \cite{kbp,VHA,ppv}).

Our approximate RG is mathematically equivalent to a domain coarsening model solved by Rutenberg and Bray \cite{rb} and Bray and Derrida \cite{bd}.  The model they solved is a limiting case of deterministic zero-temperature domain coarsening in a classical one-dimensional system with short-range interactions and a nonconserved Ising-like order parameter.  In this limit the shortest domain disappears first, allowing the two adjacent domains to grow, merge, and thus produce a new domain whose length is the sum of the lengths of the original three domains.
Our RG is also similar to those of Fisher for the ground states of certain disordered spin chains \cite{fisher}.
The difference from Fisher's RG is ``simply'' a sign: his RG can be written as
$\ell_{new}=\ell_{n-1}-\ell_n+\ell_{n+1}$, where $\ell$ in this case is the logarithm of the renormalized interaction.

\section{Critical Fixed Point Distribution}\label{sec:fixed_point_sol}

Due to the symmetry between T blocks and I blocks within our RG, the length distributions of these two types of blocks are identical at the
critical fixed point.
To derive the RG equations \cite{rb,fisher}, we define the length cutoff
\begin{equation} \label{eq:RG_Lambda_def}
\Lambda \equiv \min_n{\ell_n}
\end{equation}
and
\begin{equation} \label{eq:RG_zeta_def}
\zeta_n \equiv \ell_n - \Lambda
\end{equation}
for a block of length $\ell_n$, so $\zeta_n\geq 0$.
The RG rule (\ref{eq1}) becomes (giving the shortest block the label $n=2$)
\begin{equation}\label{eq:RG_zeta}
\zeta_{new} = \zeta_1 + \zeta_2 + \zeta_3 + 2\Lambda = \zeta_1 + \zeta_3 + 2\Lambda\,,
\end{equation}
where $\zeta_2 = 0$ because the second block is at the cutoff: $\ell_2 = \Lambda$.  Fisher's RG does not have the additive $2\Lambda$ term; it is instead $\zeta_{new}=\zeta_1+\zeta_3$.  This difference makes our fixed point rather different from Fisher's, although the same approach is used to write out the fixed point equation.  A key point in both RGs is that the length of the new block only depends on the three blocks that are removed; so if the lengths of the blocks are initially uncorrelated, no correlations are generated by this RG \cite{rb,fisher}.  And any short-range correlations between block lengths are suppressed by the coarse graining of the RG.  Thus the fixed point distribution has the block lengths uncorrelated, so we only need to study the single-block length distribution.

The probability distribution of $\zeta$ at cutoff $\Lambda$ is denoted as $\rho(\zeta, \Lambda)$.  In order to treat the critical point we now assume both types of blocks have the same length distribution.
At each RG step, when the cutoff $\Lambda$ changes by $d\Lambda$, all the blocks with $\zeta$ in the interval $[0,d\Lambda]$ are removed and new blocks are formed by combining them with their two neighboring blocks.
The distribution then needs to be renormalized and shifted back to have $\zeta_{min}=0$.
These steps all combine to produce the integrodifferential equation:
\begin{equation}
\frac{\partial\rho}{\partial\Lambda} = \frac{\partial\rho}{\partial\zeta} + \rho(0,\Lambda)\int_0^\infty d\zeta_a \int_0^\infty d\zeta_b\, \rho(\zeta_a,\Lambda)\rho(\zeta_b,\Lambda)\delta(\zeta - \zeta_a - \zeta_b - 2\Lambda)\,.
\end{equation}
As the cutoff $\Lambda$ gets large, the distribution of $\zeta$ becomes broad, and the system thus approaches an infinite-randomness fixed point.
To treat that fixed point, we divide all lengths by the cutoff, thus keeping the rescaled length cutoff at 1:
\begin{equation}\label{eq:RG_eta_definition}
\eta \equiv \frac{\zeta}{\Lambda}\,, \qquad \rho(\zeta,\Lambda) \equiv \frac{1}{\Lambda}Q(\eta,\Lambda) = \frac{1}{\Lambda}Q(\frac{\zeta}{\Lambda},\Lambda)\,.
\end{equation}
$Q(\eta)$ is then invariant under the RG flow at the critical fixed point.
We thus have the RG equation for the critical fixed point distribution $Q^*(\eta)$, independent of $\Lambda$, as
\begin{equation}\label{eq:RG_Q_star}
\frac{d}{d\eta}\Big[(1 + \eta)Q^*\Big] + Q^*(0)\Theta(\eta - 2)\int_0^{\eta - 2}Q^*(\eta_a)Q^*(\eta - \eta_a - 2)d\eta_a = 0 \,.
\end{equation}

As the total length of system is constant, $\sum_n \ell_n$ is also fixed.
At the fixed point, this becomes
\begin{equation}
\sum_i \ell_i = \Lambda \sum_i (1 + \eta_i) = \Lambda N(\Lambda) \left(1 + \langle\eta\rangle_{f.p.}\right)
\end{equation}
where $N(\Lambda)$ is the total number of blocks when the cutoff is $\Lambda$ and $\langle\eta\rangle_{f.p.}$ is the average value of $\eta$ at the fixed point, which is independent of $\Lambda$.
This implies that the product $\Lambda N$ is a constant, which results in $Q^*(0)=1/2$, and this acts as a boundary condition for Eq.\thinspace{}(\ref{eq:RG_Q_star}) so that it can be integrated out iteratively.
In the interval $0\leq\eta\leq 2$, Eq.\thinspace{}(\ref{eq:RG_Q_star}) is solved by
\begin{equation} \label{eq:RG_Q_star_solution_0_2}
Q^*(\eta) = \frac{1}{2(1+\eta)}\,, \quad \textrm{for} \quad 0\leq\eta\leq 2\,,
\end{equation}
and using this expression, in the interval $2\leq\eta\leq 4$ we have
\begin{equation} \label{eq:RG_Q_star_solution_2_4}
Q^*(\eta) = \frac{1}{1+\eta}\Big[\frac{1}{2} - \int_2^\eta \frac{\ln(\eta' - 1)}{4\eta'} d\eta'\Big]\,, \quad \textrm{for} \quad 2\leq\eta\leq 4\,.
\end{equation}

\begin{figure}[!htbp]
\begin{center}
	\includegraphics[width=0.5\textwidth]{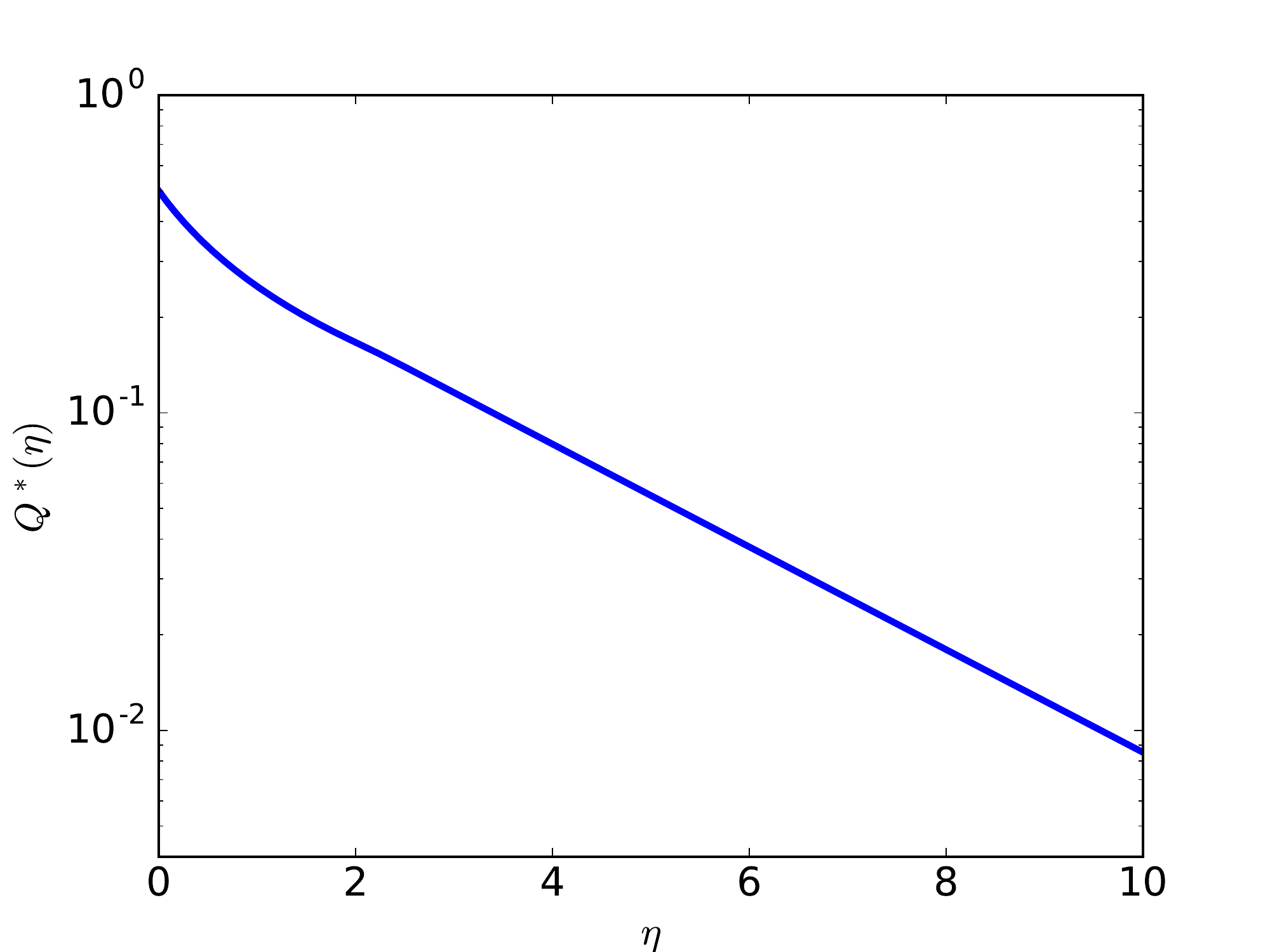}
	\caption{The scaled fixed point length distribution $Q^*(\eta)$.}
	\label{fig:RG_Q_star}
\end{center}
\end{figure}

In principle, the analytical form of $Q^*(\eta)$ for any $\eta\geq 0$ can be obtained in the same way by treating $Q^*(\eta)$ in a piecewise manner.
This shows that the physical (non-negative at all $\eta\geq 0$) solution to Eq.\thinspace{}(\ref{eq:RG_Q_star}) is unique.  The closed form solution for $Q^*(\eta)$ is shown in Rutenberg and Bray \cite{rb}.
Asymptotically $Q^*(\eta)$ falls off exponentially: $Q^*(\eta) \sim C_Q \exp(-\lambda_Q \eta)$ for $\eta \gg 1$.  The function is exhibited in Fig.~\ref{fig:RG_Q_star}.

\section{Critical Exponents}
\label{sec:critial_exponent}

By perturbing the distribution away from $Q^*(\eta)$, we can study the critical components related to the stability of this fixed point.  Moving away from the critical point generates different distributions for T and I blocks: $Q_T(\eta, \Lambda)$ and $Q_I(\eta, \Lambda)$, and when the critical fixed point is unstable under this perturbation, the difference grows as the RG flows, driving the system into either the thermal phase or the localized phase.
Similar to the derivation of $Q^*(\eta)$, the RG equations for these two distributions can be found as
\begin{equation}\label{eq:RG_Q_TI}
\left\{
\begin{split}
\Lambda\frac{\partial Q_T}{\partial \Lambda} = Q_T &+ (1 + \eta) \frac{\partial Q_T}{\partial \eta} + Q_T(\eta, \Lambda)\Big[Q_T(0,\Lambda) - Q_I(0,\Lambda)\Big] \\
&+ Q_I(0,\Lambda)\Theta(\eta - 2)\int_0^{\eta - 2} d\eta_a\, Q_T(\eta_a,\Lambda)Q_T(\eta - \eta_a - 2,\Lambda)\,; \\
\Lambda\frac{\partial Q_I}{\partial \Lambda} = Q_I &+ (1 + \eta) \frac{\partial Q_I}{\partial \eta} + Q_I(\eta, \Lambda)\Big[Q_I(0,\Lambda) - Q_T(0,\Lambda)\Big] \\
&+ Q_T(0,\Lambda)\Theta(\eta - 2)\int_0^{\eta - 2} d\eta_a\, Q_I(\eta_a,\Lambda)Q_I(\eta - \eta_a - 2,\Lambda)\,.
\end{split}
\right.
\end{equation}
To investigate the critical exponents, we consider a small perturbation away from $Q^*(\eta)$:
\begin{equation}
\left\{
\begin{split}
Q_T(\eta, \Lambda) &\equiv Q^*(\eta) + \delta_T(\eta, \Lambda)\,, \\
Q_I(\eta, \Lambda) &\equiv Q^*(\eta) + \delta_I(\eta, \Lambda)\,,
\end{split}
\right.
\end{equation}
and for $Q_{T,I}$ to still be probability distributions, $\delta_{T,I}$ both must satisfy
\begin{equation}\label{eq:RG_delta_TI_area}
\int_0^\infty \delta_{T,I}(\eta, \Lambda)d\eta = 0 ~.
\end{equation}
For simplicity in calculation, we further define
\begin{equation}
\left\{
\begin{split}
\delta_+(\eta, \Lambda) &\equiv \delta_T(\eta, \Lambda) + \delta_I(\eta, \Lambda)\,, \\
\delta_-(\eta, \Lambda) &\equiv \delta_T(\eta, \Lambda) - \delta_I(\eta, \Lambda)\,,
\end{split}
\right.
\end{equation}
so to linear order, the equations for $\delta_+$ and $\delta_-$ are
\begin{align}\label{eq:RG_delta_plus}
\Lambda\frac{\partial \delta_+}{\partial \Lambda} = \delta_+ &+ (1 + \eta) \frac{\partial \delta_+}{\partial \eta}
+ \delta_+(0,\Lambda)\Theta(\eta - 2)\int_0^{\eta - 2} d\eta_a\, Q^*(\eta_a)Q^*(\eta - \eta_a - 2) \nonumber\\
&+ \Theta(\eta - 2)\int_0^{\eta - 2} d\eta_a\, \delta_+(\eta_a, \Lambda)Q^*(\eta - \eta_a - 2)
\end{align}
and
\begin{align}\label{eq:RG_delta_minus}
\Lambda\frac{\partial \delta_-}{\partial \Lambda} = \delta_- &+ (1 + \eta) \frac{\partial \delta_-}{\partial \eta} + 2Q^*(\eta)\delta_-(0,\Lambda)
- \delta_-(0,\Lambda)\Theta(\eta - 2)\int_0^{\eta - 2} d\eta_a\, Q^*(\eta_a)Q^*(\eta - \eta_a - 2) \nonumber\\
&+ \Theta(\eta - 2)\int_0^{\eta - 2} d\eta_a\, \delta_-(\eta_a, \Lambda)Q^*(\eta - \eta_a - 2)\,.
\end{align}
To find the eigenmodes of the RG flow at the critical fixed point, we set
\begin{equation}
\left\{
\begin{split}
\delta_+(\eta, \Lambda) &\equiv \Lambda^{1/\nu_+}f_+(\eta)\,, \\
\delta_-(\eta, \Lambda) &\equiv \Lambda^{1/\nu_-}f_-(\eta)\,~.
\end{split}
\right.
\end{equation}
Under this standard scaling assumption, we have
\begin{align}\label{eq:RG_f_plus}
\frac{1}{\nu_+}f_+(\eta) = f_+(\eta) &+ (1 + \eta) \frac{d f_+}{d \eta}
+ f_+(0)\Theta(\eta - 2)\int_0^{\eta - 2} d\eta_a\, Q^*(\eta_a)Q^*(\eta - \eta_a - 2) \nonumber\\
&+ \Theta(\eta - 2)\int_0^{\eta - 2} d\eta_a\, f_+(\eta_a)Q^*(\eta - \eta_a - 2)\,.
\end{align}
and
\begin{align}\label{eq:RG_f_minus}
\frac{1}{\nu_-}f_-(\eta) = f_-(\eta) &+ (1 + \eta) \frac{d f_-}{d \eta} + 2Q^*(\eta)f_-(0)
- f_-(0)\Theta(\eta - 2)\int_0^{\eta - 2} d\eta_a\, Q^*(\eta_a)Q^*(\eta - \eta_a - 2) \nonumber\\
&+ \Theta(\eta - 2)\int_0^{\eta - 2} d\eta_a\, f_-(\eta_a)Q^*(\eta - \eta_a - 2)\,.
\end{align}
The critical exponents can then be found as eigenvalues of these two equations.
Note that the solutions also need to satisfy the normalization constraint from Eq.\thinspace{}(\ref{eq:RG_delta_TI_area}):
\begin{equation}
\int_0^\infty f_\pm(\eta)d\eta = 0 ~.
\end{equation}
On the other hand, integrating both sides of Eqs.\thinspace{}(\ref{eq:RG_f_plus}) and (\ref{eq:RG_f_minus}) gives
\begin{equation}\label{eq:RG_f_plus_minus_area}
\Big(\frac{1}{\nu_\pm} - 1\Big)\int_0^\infty f_\pm(\eta)d\eta = 0
\end{equation}
assuming both integrations exist.
Therefore the normalization constraint is automatically satisfied if the eigenvalue is not 1; otherwise we do need to check the normalization.

We diagonalized these two eigenvalue equations numerically.
The derivative was discretized to second order as a right derivative to make it well behaved even at $\eta=0$ where the functions do not exist to the left (for $\eta<0$), and the integration was discretized with the trapezoidal rule.
For $f_+$ all eigenfunctions corresponding to the largest eigenvalue of 1 are of the same sign and so are not normalizable and are thus unphysical.
The second largest eigenvalue has a real part of about $-4.4$.
This shows that $\delta_{+}$ decays at least as fast as $\sim\Lambda^{-4.4}$ when the RG is flowing and therefore is irrelevant and associated with the flow on the critical manifold toward the fixed point distribution.
This shows that the fixed point distribution $Q^*(\eta)$ is stable if $Q_T$ and $Q_I$ are perturbed in the same direction.

\begin{figure}[!htbp]
\begin{center}
	\includegraphics[width=0.5\textwidth]{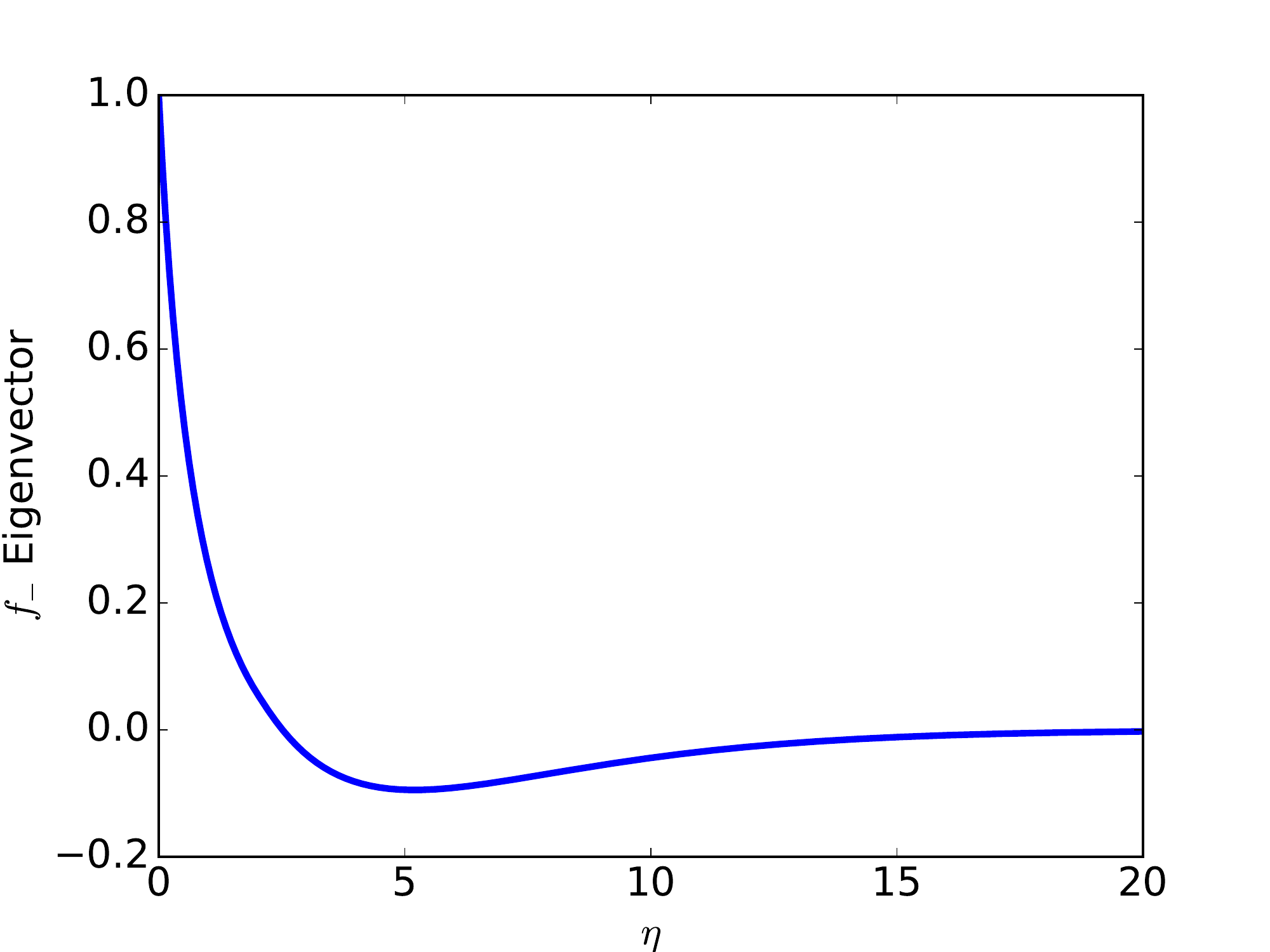}
	\caption{The one relevant eigenfunction $f_-(\eta)$ at the critical fixed point (see main text).}
	\label{fig:f_minus}
\end{center}
\end{figure}

For $f_-$, again the eigenvalue with the largest real part is $\frac{1}{\nu_-} = 1$ and its corresponding eigenfunction cannot be normalized and so is unphysical.
The eigenvalue with the second largest real part is positive and real and is the critical exponent we are looking for.  It is $1/\nu_- \cong 0.3994$, as obtained by Bray and Derrida \cite{bd}.  The next eigenvalue after that has a negative real part $\cong -1.8$, so as expected there is only one physical relevant operator at this critical fixed point.
The corresponding eigenfunction is shown in Fig.~\ref{fig:f_minus} with $f_-(0)$ normalized to $1$.
The numerical integration of it using trapezoidal rule gives about $-0.0001$, which confirms the constraint $\int_0^\infty f_-(\eta) = 0$ to the numerical precision we used.
Since this critical exponent is positive, the fixed point distribution is unstable against this perturbation which drives $Q_T$ and $Q_I$ in opposite directions and the system thus flows away from criticality into either the thermal or the MBL phase.  The difference grows as $\sim\Lambda^{1/\nu_-}$ as the length cutoff increases.

Note that this $\nu_-\cong 2.50$ satisfies and does not saturate the Chayes inequality \cite{ccfs,clo}, which in this case of $d=1$ is 
$\nu_-\geq 2$.  This is qualitatively the same as the previous RGs for the MBL transition \cite{VHA,ppv}, although this value of $\nu_-$ 
is quantitatively a little smaller than those of VHA and PVP.  On the other hand, the correlation length exponent in Fisher's RG \cite{fisher} 
does saturate the bound.  A better understanding of what determines the behavior of this exponent relative to the Chayes bound seems like 
it might be informative, but eludes us.

\section{Fractal Griffith Regions}

As the RG flows away from the critical fixed point into one of the phases, the two length distributions become very different in width.  If we flow into the thermal phase the chain becomes mostly T, with the remaining I blocks being rare and almost all of length very close to the length cutoff.  The T blocks get an arbitrarily broad length distribution, with almost all T blocks much longer than the cutoff.  Since almost all blocks of length near the cutoff are I blocks, almost all RG moves are TIT moves and make the already long T blocks even longer; our RG seems like a safe approximation for these moves.  However some small fraction of the RG moves are ITI moves with the three blocks all being of length near the cutoff.  These are the moves that maintain the population of rare I blocks (the Griffiths regions), but these are precisely the type of RG moves for which our RG approximations are not to be trusted, as we discuss above.  Thus we do not expect our approximate RG to correctly model the Griffiths regions of rare insulating segments in the thermal phase, even qualitatively.  We will return to this after examining the Griffiths regions in the MBL phase.

As the RG flows from near the critical fixed point to ``deep'' in the MBL phase, it goes into the MBL regime where almost all RG moves are ITI moves that make the already long I blocks even longer.  For these moves, our RG is typically a good approximation, because the T block that is ``integrated out'' is typically much shorter than the two adjacent I blocks, so effectively we are just joining two very long I blocks to make an even longer I block.  However, occasionally there remains a rare short I block with length at the cutoff that is slightly shorter than its adjacent T blocks, and it produces a TIT move and allows the rare T blocks to grow, thus generating the Griffiths regions, which are rare long T blocks within the MBL phase.  For these moves, our RG seems like a reasonable approximation: the two T blocks get entangled with each other across the short I block between them.

\begin{figure}[!htbp]
\begin{center}
	\includegraphics[width=0.5\textwidth]{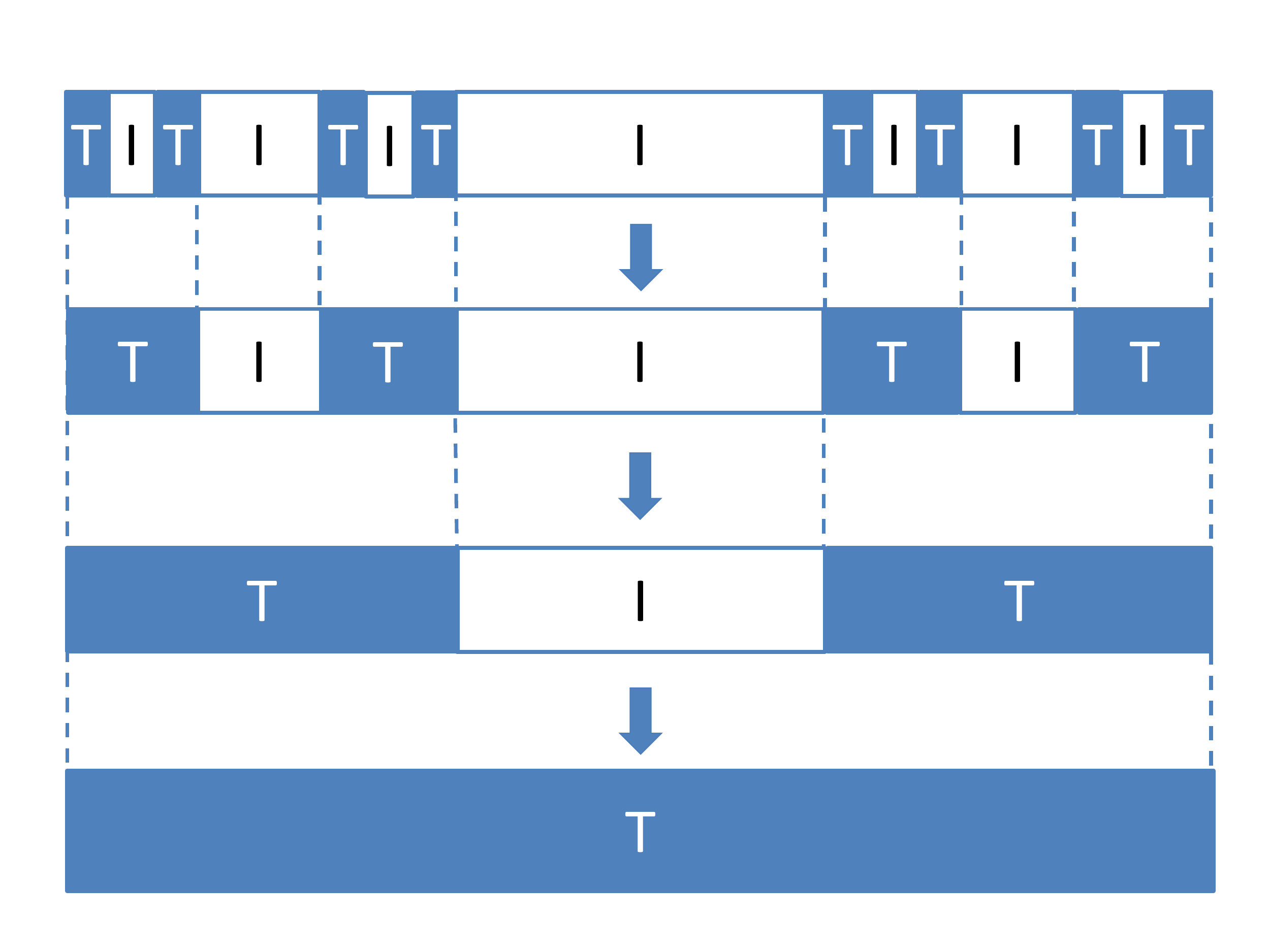}
	\caption{A sketch of part of the fractal structure of a rare large Griffiths T (locally thermalizing) region in the MBL phase.}
	\label{fig:fractal}
\end{center}
\end{figure}

Now let us ask what is the most probable way that our RG makes a large T Griffiths block within the MBL phase near the transition.  This is illustrated in Fig~\ref{fig:fractal}. The large T block arose from a TIT move.  In this limit where the RG has flowed from near the critical point to deep within the MBL phase, almost all T blocks have lengths near the cutoff. Therefore, in this TIT move the I block's length is at the cutoff and the two T blocks most likely have lengths just above the cutoff, so all three blocks are of essentially the same length.  The I block can be typical, so it typically arose much earlier in the RG by integrating out a very short T block.  The T blocks are themselves rare and each arose when the cutoff length was roughly 3 times shorter from similar TIT moves.  Thus we see that within this RG these Griffiths regions are generated from a fractal set of rare T blocks that, on scales much smaller than the final rare T block, happened to be placed just correctly such that they are able, within this approximate RG, to thermalize all the intervening typical I blocks.  This fractal set of rare T blocks has fractal dimension $d_f=\log{2}/\log{3} \cong 0.631$ in the limit of the largest such Griffiths regions.

\begin{figure}[!htbp]
\begin{center}
	\includegraphics[width=0.5\textwidth]{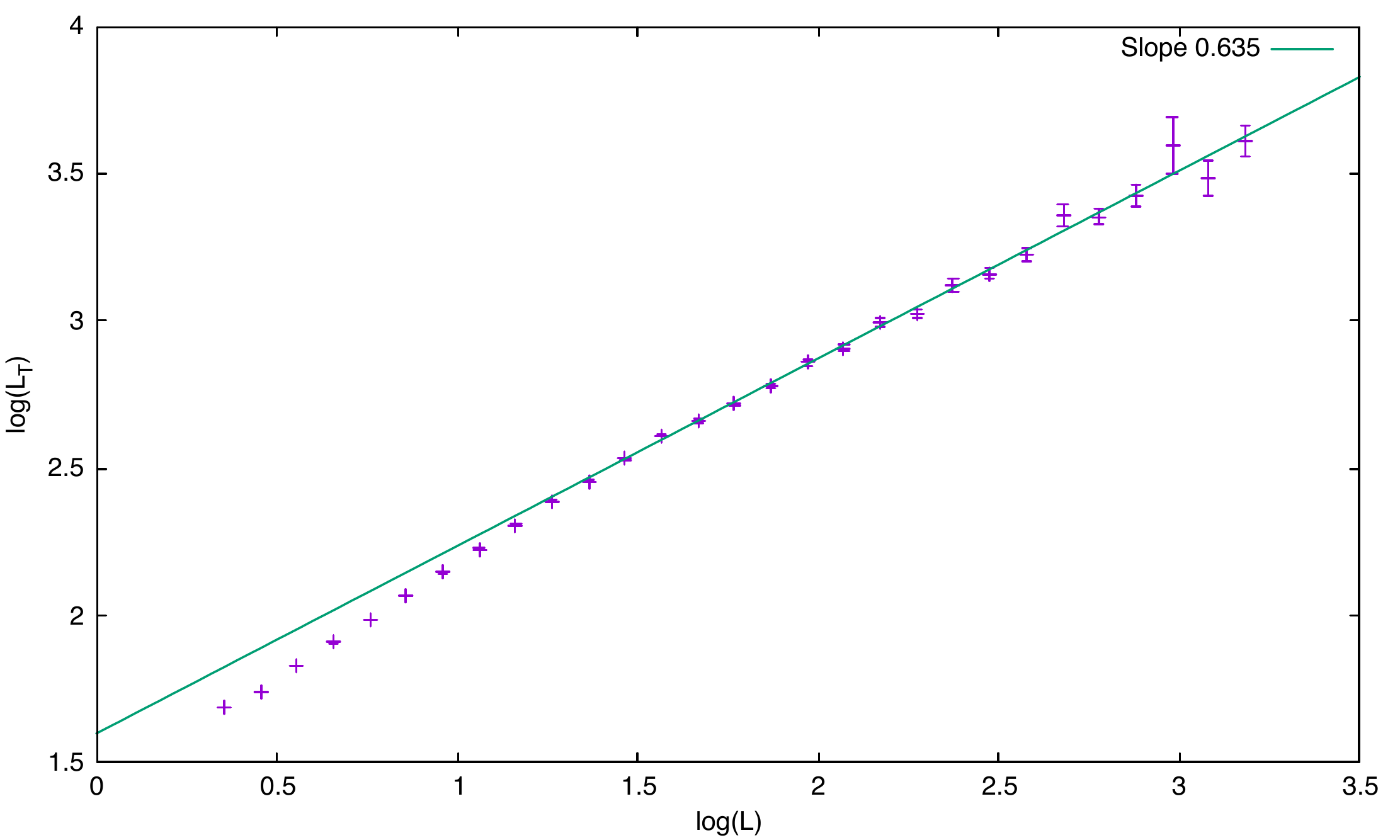}
	\caption{
		The numerical results of $L_T$ as a function of $L$ in the insulating phase to obtain the fractal dimension $d_f$. Nearby points have been binned and (their logarithm) averaged. Error bars show the standard error of points in each bin, assuming they are independent.}
	\label{fig:LT insulating}
\end{center}
\end{figure}

To test this we have numerically run the RG on a sample with initially $ 10^6 $ blocks, with
initial length distributions $Q_T(\eta) = \mathrm{e} Q^*(\mathrm{e}\eta)$ and $Q_I(\eta)= Q^*(\eta/\mathrm{e})/\mathrm{e}$, so we start near the transition and flow away into the MBL phase.  We measure $L_T$, which is the total length of each T block that is in the fractal, so it has been in T blocks at all smaller scales of the RG, and we average $L_T$ over the T blocks present when the length cutoff is $L$.
After a short transient, the results approach the expected $L_T \sim L^{d_f}$ (see Fig.~\ref{fig:LT insulating}), with
$d_f=0.635\pm0.03$, quite consistent with the expected fractal dimension.

In our simplified RG, we work only near the critical point, where the entanglement rate across an I block is close to the block's many-body level spacing.  If the bare system is actually farther into the MBL phase away from the critical point, then the entanglement rate across a typical I block will decay with block length faster than the many-body level spacing, and when we grow a fractal Griffiths region using TIT moves (in a less simplified RG such as VHA \cite{VHA}) and typical I blocks, then for the two T blocks to become entangled, the I block needs to be shorter than the two T blocks and as a result its fractal dimension $d_f$ will be larger, with $\log{2}/\log{3} < d_f <1$.  In the limit where the bare system is deep in the MBL phase $d_f$ approaches unity, and $d_f$ decreases as the transition is approached.

If this proposal of fractal Griffiths regions within the MBL phase in one dimension is not just an artifact of the approximations we make, there seem to be at least two possible scenarios:  One possibility is that the result is only partially correct: the microscopic T blocks within the fractal Griffiths region do get entangled with each other, but they do not succeed in becoming strongly entangled with, and thus locally thermalizing, the typical I blocks that are in between them within the Griffiths region.  In this case, there will be resulting correlations and entanglement within the many-body eigenstates that extend across the Griffiths regions, and so extend to distance $\ell$ with a probability that falls off as a stretched exponential function of $\ell$.  But the effective many-body level spacing of this Griffiths region may be set not by the full length of the region but only by the length $\sim\ell^{d_f}$ that is within the entangled T blocks on the microscopic scale.  This is the scenario that was mentioned and assumed in Ref.\cite{mbmott}.  The other possibility is that the Griffiths regions are fully thermal and entangled, and they respond dynamically like they have a many-body level spacing set by their full length $\ell$.  In this case the result of Ref.\cite{mbmott} is modified so the spatially averaged low-frequency conductivity $\sigma(\omega)$ behaves instead as $\log{(\sigma(\omega)}/\omega)\sim -|\log{\omega}|^{d_f}$.  Since $d_f$ is not much less than 1, for small system data this will be hard to distinguish from the continuously varying power of $\omega$ that arises\cite{mbmott} from the first scenario.  However it would be a modification to the conclusions of Ref.\cite{mbmott}, so that the Griffiths regions always dominate over the Mott many-body resonances in the limiting low-frequency conductivity of an infinite system.

Our simplified RG of course also gives fractal insulating Griffiths regions in the thermal phase, by symmetry. This is almost certainly an artifact of the oversimplifications of this RG: A low density fractal of well-placed rare insulating regions are not capable of changing an otherwise typical thermal region into an insulator.  Thus our results do not suggest that a revision of the discussions\cite{agkmd,VHA,griff} of Griffiths effects within the thermal phase is needed.

\section{Conclusions}\label{summary}

In this paper we introduced a simplified RG for the MBL phase transition in one-dimensional systems. It is mathematically equivalent to an exactly solved domain coarsening model \cite{rb,bd}, so the critical fixed point distribution and the critical exponents that characterize the stability of the critical point within our RG are known analytically or to numerical precision.
Even though some over-simplifications are incorporated to achieve this tractability, our approximate RG may correctly capture some qualitative
features of the phase transition and of the MBL phase, and might provide a basis for future more systematic RG studies.
One particular feature of this RG that we discussed in some detail is the fractal thermal Griffiths regions within the MBL phase that it produces, which seem like they might be a qualitatively correct feature of MBL in one dimension with quenched randomness.
Our RG can certainly be improved to be more realistic, but so far we do not know of a way to do so while still keeping it tractable.  Hopefully future work will be able to find some systematic way to improve these RG's.

\section{Acknowledgments}

We thank Ehud Altman, Paul Fanto, Vedika Khemani, Sid Parameswaran, Andrew Potter and Ronen Vosk for helpful discussions; Sarang Gopalakrishnan for comments on the manuscript; and Cecile Monthus for bringing Refs. \cite{rb,bd} to our attention. D.A.H. was supported by the the Addie and Harold Broitman Membership at I.A.S.

\appendix

\end{document}